# What do we learn from computer simulations of Bell experiments?

*Marian Kupczynski*

*Département de l'Informatique, Université du Québec en Outaouais (UQO), Case postale 1250, succursale Hull, Gatineau, Quebec, J8X 3X 7, Canada*

Quantum theory teaches us that measuring instruments are not passively reading predetermined values of physical observables (counterfactual definiteness). Counterfactual definiteness allows proving Bell type and GHZ inequalities. If contextuality of measurements is correctly taken into account the proofs of these inequalities may not be done. In recent computer simulations of idealized Bell experiment predetermined successive outcomes of measurements for each setting and predetermined time delays of their registrations are calculated. Time–windows and time delays are used to post- select various sub-samples. Correlations, estimated using these post-selected samples are consistent with the predictions of quantum theory and the time-window dependence is similar to the dependence observed in some real experiments. It is an important example how correlations could be explained without evoking quantum nonlocality. However by using a suitable post-selection one may prove anything. Since before the post-selection generated samples may not violate CHSH as significantly as finite samples generated using quantum predictions thus one may not conclude that counterfactual definiteness is not able to distinguish classical from quantum physics. Moreover we show that for each choice of a time-window there exists a contextual hidden variable probabilistic model consistent with the post-selection procedure used by the authors what explains why they were able to reproduce quantum predictions.

**I. Introduction**

In recent papers Hans de Raedt et al. [1, 2] present results of computer simulations of idealized EPRB spin polarization correlation experiment . For each member of a "photon pair" predetermined outcomes of measurements (counterfactual definiteness) and a predetermined time delays of outcomes registration are calculated.

The authors use a particular ingenious choice of time delays and post-select various sub-samples of data in strongly setting and time–window dependent way. Correlations, estimated using these post-selected samples, agree remarkably with the correlations predicted by quantum theory (QT). Moreover the time -window dependence of the estimates mimics the dependence observed in real experiment [3].

Since outcomes and time delays are created in a locally causal way the simulations provide a nice example that one does not necessarily need to evoke quantum nonlocality in order to explain the violation of Bell inequalities.

However the time delays and post-selected samples are strongly setting dependent. <u>If no post-selection is used generated samples may not violate CHSH inequalities as significantly as simulations based on quantum probabilistic model</u>. This is a well-known fact e.g. [4, 5]. It is also well known that by using setting dependent post-selection one may prove anything [6-9]. Moreover the time delays and a particular post-selection procedure, used by the authors, may not explain the violation of Bell-type inequalities in all recent experiments [10-16].

Therefore we disagree with the conclusion that CFD, <u>as it is usually defined</u>, *does not distinguish classical from quantum physics* [1].

The paper is organized as follows. In section 2 we recall the definition of *counterfactual definiteness.* In section 3 we discuss details and results of computer simulations before post-selection. In section 4 we comment on the post-section procedure used in [1.2]. In section 5 we explain what do we mean by saying that QT is a contextual theory and we clarify the notion of *contextuality loophole*. In section 6 we discuss how computer simulations may contribute to testing whether quantum theory is predictably complete. In section 7 we list some conclusions.

## II. Counterfactual definiteness

At first let us recall a generally accepted definition of *counterfactual definiteness* (CFD): *values of all observables are predetermined before a measurement and recorded passively by measuring instruments*. The existence of incompatible quantum observables clearly proves that various instruments are playing an active role during measurements thus CFD is inconsistent with QT [17-22]. The lattices of classical and quantum observables are incompatible [23-26].

According to CFD " photon pairs" in spin polarization correlation experiments (SPCE) are described by predetermined values of spin projections in all directions registered passively by measuring instruments. Using CFD one may define joint probability distributions on the same probability space and one may deduce, using it, the correlations between spin projections for all different settings used in SPCE. CFD is also crucial in all finite- sample proofs of Bell- type inequalities [22, 26-55]. The list of references is by no means complete.

To define local realistic hidden variable model <u>one does not need to assume that various spin projections may be measured at the same time in all directions on a given pair of particles</u>. One assumes only that there exists a probability distribution of some hidden variables λ from which all probabilistic predictions for feasible pairs of experiments may be deduced. As we explained in detail in [26] local realistic hidden variable models are not Kolmogorov probabilistic models but they are isomorphic to particular Kolmogorov models in which joint probability distributions of all possible values of spin projections are well defined.

## III. Simulation protocols and results

Two different computer models used in [1, 2] generate some finite samples consistent with the existence of the joint probability distribution discussed above.

The authors , using a <u>different</u> definition of CFD, say that one of their computer experiments is run according to a CFD-compliant protocol and another is run according to a non-CFD-compliant protocol. In both models a state of a "pair" is described by ($\varphi$, $r_1$; $\varphi + \pi/2$, $r_2$) and for this state all the outcomes of "measurements" are predetermined. Two lines below explain how an outcome x and a time delay t* are calculated for each member of a "pair" passing by a polarizer described by ($a$, T) where T is a fixed parameter related to a time unit and $a$ is an angle:

$$1: \text{compute } c = \cos[2(a - \varphi)]; \; s = \sin[2(a - \varphi)]; \quad (1)$$
$$2: \text{set } x = \text{sign}(c); \; t^* = rTs^2 \quad (2)$$

where $\varphi$ is randomly drawn from [0,$2\pi$] and $r$ from [0,1].

.

According to the definition of CFD, given by us in the preceding section, <u>both protocols are CFD- compliant. This is why we call them: Protocol 1 and Protocol 2.</u>

1. Protocol 1 (non- CFD-compliant) <u>is consistent with a realisable protocol in SPCE</u> since outcomes, for any pair of settings , are generated each time for a different "photon pair". Namely 4 pseudo- random time series of data for 4 different pairs of settings are generated:

$$(x_1(t), x_2(t), t_1^*(t), t_2^*(t)) = G(\varphi_1(t), \varphi_2(t), r_1(t), r_2(t), a_1(t), a_2(t)) \quad (3)$$

where t=1,…4N and $a_1(t)\, a_2(t)= a_1\, a_2$ for t=1…N; $a_1(t)\, a_2(t)= a_1\, a'_2$ for t=N+1….2N ; $a_1(t)\, a_2(t)= a'_1\, a_2$ for t=2N+1…3N; $a_1(t)\, a_2(t)= a'_1\, a'_2$ for t=3N+1….4N .

2. Protocol 2 (CFD-compliant<u>) is impossible to realize in SPCE and in QT</u> because it calculates 4 predetermined values of a simultaneous joint measurement of incompatible observables performed on each "photon pair". This pseudo-random time series can be defined as:

$$(x_1(t), x'_1(t), x_2(t), x'_2(t), t_1^*(t), t_1'^*(t), t_2^*(t), t_2'^*(t)) = G_1(\varphi_1(t), \varphi_2(t), r_1(t), r_2(t), a_1, a'_1, a_2, a'_2) \quad (4)$$

where t=1,…4N

Richard Gill [4] studies an impact of CFD on samples which are created in idealized SPCE experiment with random choice of the settings. He defines a 4 x 4N spreadsheet. Each line of this spreadsheet describes each incoming "photon pair" by predetermined values ±1 of spin projections for 4 settings (a, a, b, b') used in SPCE. If no constraints are imposed one can have only 16 <u>different</u> lines in the spreadsheet which are permuted depending on a sequence of " photon pairs" arriving to detectors.

In SPCE pairs of settings are chosen randomly and only the results for one pair of settings can be observed each time. Therefore Gill constructs his samples by choosing from each line two outcomes corresponding to the randomly chosen settings. Because of the predetermination of outcomes (CFD) the finite samples created following this protocol may not violate CHSH as significantly as predicted by QT and Gill makes a conjecture:

$$\Pr\left(\langle AB\rangle_{obs} + \langle AB'\rangle_{obs} + \langle A'B\rangle_{obs} - \langle A'B'\rangle_{obs} \geq 2\right) \leq \frac{1}{2} \qquad (5)$$

The samples containing the outcomes $x_1, x'_1, x_2$ and $x'_2$ created using (1-4) have similar properties as samples extracted from the counterfactual 4 x 4N spreadsheet studied and discussed in [4, 22]. Namely:

- Protocol 1 chooses, for each "photon pair "only two entries from a corresponding line of the spreadsheet: one for Alice and one for Bob. It is obvious from (3) that the first N pairs of entries are chosen from the same pair of columns since the settings are changed systematically only at t=N+1, 2N+1, 3N+1, 4N+1. In spite of the fact that pairs of settings are not chosen randomly generated samples have the same properties as the samples studied by Gill.

- No matter how settings are chosen, for each "photon pair", a complete line of the spreadsheet is outputted by Protocol 2. A produced spreadsheet is a sample drawn from some CFD -compliant joint probability distribution of 4 random variables. It is easy to see [4, 22] that in this case |S|=2 for any finite sample thus CHSH inequality [56, 57] is never violated.

For Protocol 1: results of computer simulations, before a post- selection, are consistent with Gill's conjecture (54 samples out of 100 violate |S| ≤2). For Protocol 2 : CHSH is never violated. Thus <u>computer simulations confirm that CFD is not consistent with QT.</u>

### IV. Setting–dependent post-selection

How can we understand that post- selected samples for both protocols comply nicely with quantum predictions? The authors post- select samples in strongly setting- dependent, way. <u>It is not a fair sampling</u>. Time delays t* are outputted for each pair. As in Weihs et al. [3] differences of time-delays and time-windows are used to post-select final samples. One has to admit that the agreement with QT predictions and similarity to the experimental results of Weihs et al. is remarkable.

Authors acknowledge that: models that incorporate post selection and/or exploit so called coincidence loophole are known to produce results [6-9] which may violate $|S| \leq 2$. We will add a simple example showing that a post-selection not based on a fair sampling leads to inconsistent results and conclusions.

We want to analyze impact of a post-selection on two perfect simple random samples $S_1=\{x_1,\ldots x_n\}$ and $S_2=\{y_1,\ldots y_2\}$ where $x_i=\pm 1$ and, $y_i=\pm 1$. Using a simple pairing of outcomes we obtain a sample $S_3=\{(x_1, y_1),\ldots (x_n, y_n)\}$.

- If we post-select using a criterion *keep only if $x_i + y_i = 2$* we get completely correlated sub-sample
- If we post-select using a criterion *keep only if $x_i + y_i = -2$* we get strogly anti-correlated sub-sample
- If we post-select using a criterion *keep only if $x_i + y_i = 0$* the correlations vanish.

Authors say that the model using Protocol 1 *suffers from contextuality loophole*. We find that this statement may be confusing. Therefore In the next section we explain what is meant usually by a *contextuality loophole*.

### V. The meaning of contextuality loophole

*Contextuality* has a different meaning for different authors [20, 26, 44, 49, 54, 58, 62]. For some authors a theory is non-contextual if values of physical observables do not depend on a specific experimental protocol used to measure these observables. For example a length of a table does not depend how this length is measured. Similarly QT does not say how a linear momentum of an electron and its energy have to be measured. Quantum state vector represents all equivalent preparation procedures and a self-adjoint operator represents all equivalent measurement procedures of a corresponding physical variable [62]. Therefore one can find statements in the literature that as well the classical physics and QT are non-contextual theories [60].

Physical systems may have different properties. Some of them are attributive what means that they do not change in the interaction with measuring instruments and they do not depend on the environment in which they are measured. Attributive properties are for example: length of a table (in my house), inertial mass, electric charge of an electron, etc. Physical systems may be also described by contextual properties which are only revealed in particular experimental and/or environmental contexts for example: color, weight, magnetization, spin projection etc.

Interesting contextual property is a probability (understood not as a subjective belief of a human agent) which is neither a property of a coin nor a property of a flipping device. It is only a property of a whole random experiment [33, 26, 44, 49]

In quantum theory, as Bohr insisted, we deal with: "the impossibility of any sharp distinction between the behavior of atomic objects and the interaction with the measuring instruments which serve to define the conditions under which the phenomena appear" [61, 63, 64].

In particular QT gives probabilistic predictions on a statistical scatter of outcomes obtained in repeated "measurements" of some physical observable on identically prepared physical systems. Since the probability is only a property of a random experiment performed in a particular experimental context thus QT provides a contextual description of the physical reality. Similarly Kolmogorov probabilistic models provide contextual description of random experiments. Namely each random experiment (in which there is a statistical stabilization) is described by its Kolmogorov model defined on a dedicated probability space [35, 44].

Probabilistic models as well as QT do not enter into details how individual outcomes are obtained but provide only predictions for the experiments as a whole. If the context of an experiment changes (for example we have two slits open instead of one) then Kolmogorov and quantum probabilistic models change. This is why we say : QT is a contextual theory. Only rarely different random experiments may be described using marginal probabilities deduced for some joint probability distribution [65] and it is obvious that the joint probability distribution for incompatible quantum observables do not exist.

Of course to describe a detailed time -dependence of random phenomenon we may not use simple probabilistic models but only Stochastic Processes.

After this long introduction let us clarify the notion of *contextuality loophole* (called like this for the first time by Theo Nieuwenhuizen [54] ). We say that a probabilistic hidden variable model suffers from *contextuality loophole* if it uses the same probability space to describe different incompatible random experiments. To close *contextuality loophole* a model has to incorporate supplementary parameters describing measuring devices [66]. If setting dependent parameters are correctly incorporated in a hidden variable probabilistic model then Bell-type inequalities may not be proven.

For example if instead of (1,2) we define $x_a(t)=f_a(\lambda_1(t), \lambda_a(t),a)$ and $x_b(t)=f_b(\lambda_2(t), \lambda_b(t),b)$ where $(\lambda_1(t), \lambda_2(t))$ and $(\lambda_a(t), \lambda_b(t))$ describe respectively an "EPR-pair" and "microstates" of measuring instruments (a, b) in the moment of the" measurement" then Gill's counterfactual spreadsheet does not exist and the only constraint we have is $|S| \leq 4$. The same constraint was obtained Andrei Khrennikov in his Kolmogorov model for SPCE experiments [46, 47].

Therefore any hidden variable model wanting to reproduce QT predictions must include explicit dependence of hidden variables on the settings. This dependence on the settings does not imply nonlocal interactions. For example Zhao et al. [59] generated samples consistent with QT predictions for EPRB experiment. They also derived a family of hidden variable models which were consistent with their CFD-compliant generation of data and with their setting- and window-dependent post-selections. These probabilistic models are in fact contextual. We rewrite here Eq. 19 from [67] in explicitly contextual form:

$$P(x_1, x_2 | \alpha_1, \beta_1, W) = \int_0^{2\pi} \int_0^{2\pi} P(x_1 | \alpha, \xi_1) P(x_2 | \beta, \xi_2) P(\xi_1, \xi_2 | \alpha, \beta, W) \, d\xi_1 \, d\xi_2 \qquad (6)$$

It is an interesting example how a particular post-selection induces *contextuality* in a random experiment. Contextual probabilistic models of spin polarization correlation experiments may also be defined in more direct and intuitive way see for example [22, 26].

## VI. Computer simulations and predictable completeness of QT

Bohr claimed that any subquantum analysis of quantum phenomena is impossible and that QT provides a complete description of individual physical systems [17]. Einstein believed that quantum probabilities are emergent and that a complete theory should give more details how individual outcomes are created [63, 68]. Bohmian- mechanics, stochastic electrodynamics, hidden variable models and various computer simulations are attempts to realize this program in particular cases.

Event-by-event simulations of quantum experiments are trying to get an intuitive understanding of quantum phenomena without evoking quantum magic. Several quantum phenomena were simulated recently with success e.g. [69-71]. These simulations have no ambition of replacing QT. They generate time series of outcomes similar to those created in real experiments and study the effects which QT is unable to address. As we demonstrated in a recent paper [72] <u>various experimental protocols, based on the same probabilistic model, provide more detailed information how data are created and may produce significantly different finite samples</u>.

For example in [3] one has to use different time- windows in order to identify correlated detection events. The estimated correlations strongly depend on time- windows used and this dependence was studied. QT has nothing to say about these time-windows in contrast to computer simulations discussed above, which are able to reproduce time- window dependence of correlations observed in the real experiment.

The main ingredient of the computer model discussed above are time delays. The authors say <u>only</u> that these time delays are due to the existence of dynamic *many-body interactions of the photon* with measuring apparatus. In our opinion the equations (1) and (2) allow to imagine more detailed mechanism how data in real experiment might be created.

First of all let us change slightly the assignment of the variables. Let us describe an "EPR-pair" by $(\varphi, \varphi + \pi/2)$ and measuring devices by $(\alpha, r_1, T; \beta, r_2, T)$. When entering linear polarizers "magnetic moments" of each pair are pointing in opposite directions. They are aligned along the directions $a_1$, and $a_2$ respectively and they are sent to corresponding detectors for a registration. It is plausible that the average time needed for this alignment increases and decreases in the function of ($a$- φ). The model would have been closer to this physical intuition with parameters *r*'s drawn randomly from [1-*c*, 1] (where *c* is a small number) instead from [0,1].

Probably such reformulation would not change significantly the simulation results. After this reformulation time delays are no longer predetermined by a value of φ thus CFD does not hold for time delays.

Of course the dependence of time delays on ($a$- φ) in (2) was chosen to reproduce QT predictions. Nevertheless the existence of conjectured time delays, not predicted by QT, may be tested in appropriate experiments with polarized beams. The possibility of the existence and implications of time delays in Bell experiments were discussed, for the first time in detail, by Saverio Pascazio [7]. We don't know whether such delays were studied experimentally. If such time delays were discovered it would give another argument in favor of the idea that QT is not predictably complete.

In several papers we advocated that in order to check the predictable completeness of QT one has to search for unexpected regularities in the experimental time-series of data [73-75].

### VII. Conclusions

If we had only the data of Weihs et al. [3] we could find the computer model discussed above plausible. However the violation of Bell–type inequalities was demonstrated in many experiments following completely different experimental protocols e.g. [10-16]. In some of them there is no place for the post-selections based on time-delays, used in [1, 2, 67], which were needed in order to reproduce QT predictions.

Besides CFD is clearly incompatible with predictions of QT for successive polarization measurements [19] and there is also no place ,in these experiments, for setting- dependent time delays and time-windows.

Therefore, in our opinion, the results of [1, 2. 67] do not allow to make a general statement that: *CFD does not separate or distinguish classical from quantum physics.*

CFD cannot be maintained in QT because quantum observables are contextual what means that their values are not predetermined and that "measuring" instruments participate actively in the creation of the results obtained in well-defined experimental contexts. As Peres told: *unperformed experiments have no results* [76].

The *contextuality* as we define it, is not restricted to quantum phenomena [69]. Many experiments in other domain of science may not be described using a common probability space and the violation of Bell-type inequalities may be observed e.g. [44, 45, 77]. One may even perform particular experiments with colliding balls in classical mechanics such that the outcomes violate one of Bell inequalities [51].

Let us now list what we learned from the computer simulations of Bell experiments [1,2,67]:

1. To explain the correlations in EPRB one does not need to evoke quantum nonlocality.

2. Generated finite samples, before post-selection, cannot violate CHSH as significantly as it is predicted by QT.

3. One has to be careful when applying a post-selection procedures and time-windows in real experiments. Using unfair post-selection one may prove anything.

4. The computer model presented in the paper is able to reproduce the predictions of QT only because in corresponding probabilistic models hidden variables depend explicitly on the settings (4). The *contextuality* of these models is induced by setting dependent post-selections.

5. The time delays introduced by the authors may suggest that a time of a passage across a linear polarizer may depend on a polarization of the entering beam. This conjecture may be tested experimentally.

6. Computer simulations allow more detailed study of created time series of data than it is possible in QT and may inspire more detailed study of real data in order to check whether QT is predictably complete.